\documentclass{llncs}

\usepackage{algo}  
\usepackage{epsfig}  
\usepackage{url}  
\usepackage{latexsym}   
\usepackage{boxedminipage}  
\begin{document}

\newcounter{algo}  
\setcounter{algo}{0}  
\newenvironment{algo}[1]{%  
\refstepcounter{algo}  
%\begin{figure}[htbp]  
\begin{center}\sl Algorithm \thealgo: #1  
\begin{boxedminipage}{0.95\textwidth}  
}{\end{boxedminipage}\end{center}}%\end{figure}}  

\newcommand{\mC}{\mathcal{C}}  
\newcommand{\mF}{\mathcal{F}}  

\newenvironment{preuve}{\noindent {\it Proof.}}{$\Box$\vskip1ex}  
\newenvironment{preuveA1}{\noindent {\it Proof of conservation of   
Property $A_1$.}}{$\Box$\vskip1ex}  
\newenvironment{preuveA2}{\noindent {\it Proof of conservation of   
Property $A_2$.}}{$\Box$\vskip1ex}  
\newenvironment{preuveB}{\noindent {\it Proof of the invariants.}}  
{$\Box$\vskip1ex}  
\newenvironment{OJO}{\noindent {\bf OJO:}}{$\Box$\vskip1ex}

\title{Consecutive ones property testing: cut or swap}

\author{
Mathieu Raffinot\inst{1}}
\institute{LIAFA, Univ. Paris Diderot - Paris 7, 75205 Paris Cedex 13, France.\\
{\tt raffinot@liafa.jussieu.fr}}
\maketitle

%%%%%%%%%%%%%%%%%%%%%%%%%%%%%%%%%%%%%%%%%%%%%%%%%%%%%%%%5  
\begin{abstract}
Let ${\cal C}$ be a finite set of $n$ elements and ${\cal
R}=\{R_1,R_2, \ldots , R_m\}$ a family of $m$ subsets of ${\cal
C}$. The family ${\cal R}$ verifies the consecutive ones property if
there exists a permutation $P$ of ${\cal C}$ such that each $R_i$ in
${\cal R}$ is an interval of $P$. There already exist several
algorithms to test this property in $O(\sum_{i=1}^m |R_i|)$ time,
all being involved. We present a simpler algorithm, based on a new
partitioning scheme.
\end{abstract}  
  
%%%%%%%%%%%%%%%%%%%%%%%%%%%%%%%%%%%%%%%%%%%%%%%%%%%%%%%%5  
\section{Introduction}  
\label{sec:intro}

\begin{figure}[htb]
\vspace{-0.3cm}
  \centering
\includegraphics[width=12cm]{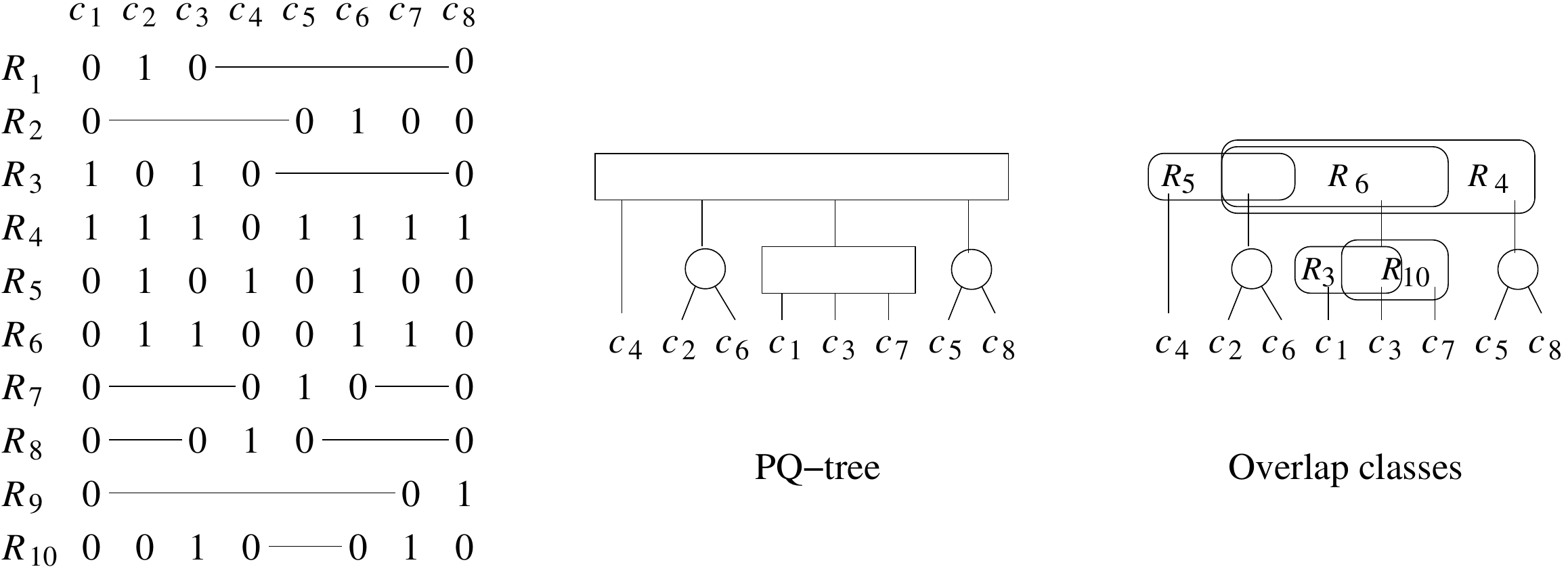}
\caption{A matrix verifying the consecutive ones property, its associated PQ-tree and the information contained in overlap classes. In the PQ-tree, $Q$ nodes are represented by boxes, while $P$ nodes by circles.}
 \label{matrice}
\vspace{-0.3cm}
\end{figure}

Let ${\cal C} = \{c_1, \ldots ,c_n\}$ be a finite set of $n$ elements
and ${\cal R}=\{R_1,R_2, \ldots , R_m\}$ a family of $m$ subsets of
${\cal C}$. Those sets can be seen as a 0-1 matrix, where the ${\cal
C}$ represents the columns and each $R_i$ the ones of row $i$. Figure \ref{matrice} shows such a matrix.

The family ${\cal R}$ verifies the consecutive ones property (C1P) if
there exists a permutation $P$ of ${\cal C}$ such that each $R_i$ in
${\cal R}$ is an interval of $P$. For instance, the family given by
the matrix of \ref{matrice} verifies C1P. Efficiently testing C1P has
received a lot of attention in the literature for this problem to be
strongly related to the recognition of interval graphs, the
recognition of planar graphs, modular decomposition and others graph
decompositions. The consecutive ones property is the core or many
other algorithms that have applications in a wide range of domains,
from VLSI circuit conception through planar embeddings
\cite{nr-pgd-04} to computational biology for the reconstruction of a
chromosome from a set of contigs \cite{ChKe99}. We denote $|{\cal R}|
= \sum_{i=1}^m |R_i|$. Several $O(|{\cal R}|)$ time algorithms have
been proposed to test this property, following five main approaches.

The first approach and still the most well known one is the use of
PQ-tree structure \cite{BL76}. A PQ-tree is a tree that represents a
set of permutations defined by the possible orders of its leaves
obtained by changing the order of the children of any internal node
depending of its type which can be $P$ or $Q$. For a $P$ node, any
order of its children is valid, while for a $Q$ node only the complete
reversal of its children is accepted. For instance, in Figure
\ref{matrice}, the PQ-tree represents the order $c_4 c_2 c_6 c_1 c_3
c_7 c_5 c_8$, but also $c_4 c_2 c_6 c_7 c_3 c_1 c_5 c_8,$ $c_4 c_6 c_2
c_7 c_3 c_1 c_8 c_5,$ and so on. The main point for using PQ-trees is
that if a family verifies C1P, then one can build a PQ-tree
representing exactly all column orders for which the C1P will be verified. For
instance, the PQ-tree in Figure \ref{matrice} represents all orders
for which the family given by the matrix at its right verifies
C1P. If a family does not verify C1P, its associated PQ-tree is said
empty.

Given a family, in order to build its associated PQ-tree, each row is
inserted one after the other in the tree while the PQ-tree is not
empty. This update is done through a procedure called {\tt Refine}
which complexity is amortized on the size of the tree. The main
drawback of this approach is that the implementation of {\tt Refine}
in its linear time complexity is still a challenge. It uses a series of 11
templates depending on the form of the tree and choosing which to use
in constant time is a huge programming difficulty, that has only
slightly been reduced by Young \cite{Young1977} using a recursive {\tt
Refine} that allows us to reduce the number of templates. Moreover,
extracting a certificate that the family does really not verify C1P
from this approach is hard. Therefore, given a PQ-tree
implementation, one can hardly be confident neither in its validity nor in
its time complexity. This is the reason why many other algorithmic
approaches have been tempted to test C1P using simpler and/or
certified algorithms.

One of those attempt consists in first transforming the C1P testing
problem to interval graph recognition by adding fake rows and then use
a special LexBFS traversal that produces a first order on $C$ that has
some special properties \cite{HMcCPV00}. A recursive partitioning
phase is then necessary following both this LexBFS order and an order on
the rows derived from a clique tree built from the LexBFS
traversal. This approach is also complex, both to understand and to
program, and surprisingly the links between these two first approaches
are not that clear.

A third approach was to try to design the PC-tree \cite{HMC03}, an
easiest structure to refine than the PQ-tree. However as Haeupler and
Tarjan noticed in \cite{HaeuplerT08}, the authors of \cite{HMC03} did
not consider "implementations issues" (sic) than lead to incorrect
algorithms for C1P testing and planar graph recognition.

A fourth approach appeared in \cite{Hsu2002} with the idea of
simplifying the C1P test by avoiding PQ-tree. However, the algorithm
remains very involved.

A last and more recent approach has been presented by R. McConnell in
\cite{McConnell04}. This approach is a breakthrough in the
understanding of the intrinsic constraints of C1P and the real nature
of the PQ-tree. We describe this approach in details since our method
is a tricky simplification of it. McConnell shows that each $Q$ node
of the PQ-tree represents in fact an overlap class of a subset of the
rows. Two rows $R_i$ and $R_j$ of ${\cal R}$ overlap if $R_i \cap R_j
\neq \emptyset,$ $R_i \setminus R_j \neq \emptyset,$ and $R_j
\setminus R_i \neq \emptyset.$ An overlap class is a equivalence class
of the overlap relation, that is, two rows $R_i$ and $R_j$ are in the
same class if there is a chain of overlaps from $R_i$ to $R_j$.  For
instance, the two non trivial overlap classes of the family example
given by the matrix of Figure \ref{matrice} are shown on the same figure on
the right. Overlap classes partition the set of rows and form a
laminar family, and thus they can be organized in an inclusion tree.

This tree is the skeleton of the PQ-tree and the remaining $P$ node
might also been derived from the overlap classes. However, for an
equivalence class to be a node of the PQ-tree, it also has to verify
the consecutive one property. Thus, where is the gain ? The trick used
by McConnell is that verifying the C1P of an overlap class is
independent of the other overlap classes and somehow easier provided a
spanning tree of the overlap graph of the class. Using a partitioning
approach guided by this tree, it is linear in the total size of the
rows in an overlap class to test if this overlap class verifies
C1P. Consequently, by testing overlap classes one after the other, one
can verify if the whole set ${\cal R}$ fulfills C1P in $O(|{\cal R}|)$
time. The technical complexity of the approach is twofold: (a) compute
overlap classes and (b) a spanning tree of each class.

Point (a) is performed in \cite{McConnell04} through an algorithm of
Dahlhaus published as a routine of \cite{Dahlhaus00} used for undirected graph
split decomposition. It is considered by McConnell as a black box
that takes as input ${\cal R}$ and returns a list of overlap classes and
for each overlap class the list of rows that belongs to.

Point (b) is then computed in \cite{McConnell04} for each overlap
class by a complex add-on from the list of rows in the class.

In this article we present a simplification of this last approach by
introducing a new partitioning scheme. It should be noted first that
McConnell's approach can already be very slightly simplified using existing
tools. Indeed, the algorithm of Dahlhaus for computing overlap classes
is an algorithmic pearl that has been recently simplified and made
computationable in the sense that the original version uses an LCA
while the simplified version presented in \cite{CharbitHLMRR08} only
uses partitioning. Moreover, a modification of Dahlhaus's approach
allows us to extract a spanning tree of each overlap class. This
modification is not obvious but remains simpler than the add-on of
\cite{McConnell04}. However, building a spanning tree from Dahlhaus is
intrinsically difficult, because the two concepts are somehow antinomic:
Dahlhaus's approach maintains some ambiguities in the row overlaps that
permit to gain on the overall computation, while computing a spanning
tree requires solving most of these ambiguities, which is sometimes
difficult. In this paper, we successfully maintain these ambiguities
even in the partitioning phase, avoiding buliding a spanning tree.

To clearly present our approach let us consider the difference between
the PQ-tree approach and that of McConnell in terms of
partitioning. The PQ-tree records a partition of $\cal{C}$ induced by
the rows even if some rows can be included in others (a row might not
{\em cut} any class of the partition). The difficulty arises when
updating the structure: in the same time we need to update both a
partition and an inclusion tree that are intrinsically merged. In the
second approach the idea is to impose that each row added surely
overlaps a previous one, which simplifies the partitioning since the
inclusion tree as not to be maintained. This also insures the linear
time complexity without any amortizing need, but at the cost of the
computation of a spanning tree of each overlap class.

Our approach lies in between. For each overlap call we update a
partition, but we also allow some fail and swap in the partitioning
scheme. We compute an order that guaranties that when adding a new row
$R_1$, if it does not overlap any row already considered, then the
row following $R_2$ will, and moreover $R_1$ overlaps $R_2$ and will
be considered next. We thus swap $R_1$ and $R_2$ in the order we
update the partition if $R_1$ does not cut.  We call this order a
"swap overlap order". This order could of course be obtained from a
spanning tree, but we explain below how we can compute such an order
at a very small computational price by entering deeper in Dahlhaus's
algorithm, that we also slightly simplify for our needs. Our algorithm
thus runs in 3 main steps: (1) the computation of each overlap class
using an algorithm close to that of Dahlhaus, (2) for each class we
compute of a swap overlap order, and (3) we partition each
class guided by this order using a new partitioning scheme. If the
partitioning fails on a class, the C1P is not verified. Steps 1 and 2
are performed in the same time, but for clarity we present them in two
distinct steps.

This article is organized as follows. In the following Section
\ref{overlap} we present two variations of Dahlhaus's algorithm for
computing overlap classes. In Section \ref{swap} we explain our main
notion of {\em swap overlap order} and explain how to slightly modify
Dahlhaus's algorithm to generate such an order for each overlap
class. In Section \ref{partition-refine} we eventually explain how to
test C1P on each overlap class using the swap overlap order associated
to.  We added two appendixes. The first is an example of the
construction of a swap overlap order. The second is a technical
routine used in Dahlhaus's algorithm revisited in \cite{CharbitHLMRR08}
that we mainly recall.
 
\section{Computing Overlap Classes}
\label{overlap}

In this section we recall and slightly modify the algorithm of
Dahlhaus for computing overlap classes already simplified and
presented in \cite{CharbitHLMRR08}. The computational problem to efficiently
compute the overlap classes comes from the fact that the underlying
overlap graph, where $R_i$ are the vertices and $(R_i,R_j)$ is an edge
if $R_i$ overlaps $R_j$, might have $\Theta(|{\cal R}|^2)$ edges and thus be
quadratic in $O(|{\cal R}|).$ An {\em overlap class} is a connected
component of this graph.

Let $\mbox{LR}$ be the list of all $R \in {\cal R}$ sorted in
decreasing size order. The ordering of sets of equal size is
arbitrarily fixed, and thus $\mbox{LR}$ is a total order. Given $R \in {\cal
R}$, we denote $\mbox{Max}(R)$ as the largest row $X \in {\cal R}$
taken in $LR$ order such that $X <_{LR} R$ and $X$ overlaps
$R$. This definition is modified from that in \cite{CharbitHLMRR08} to consider
the order $\mbox{LR}$ in the definition of $\mbox{Max}(R).$

Note that $\mbox{Max}(R)$ might be undefined for some sets of ${\cal
R}.$ In this latter case, in order to simplify the presentation of
some technical points, we write $\mbox{Max}(R)=\emptyset.$ Dahlhaus's
algorithm is based on the following observation:

\begin{lemma}[\cite{Dahlhaus00,CharbitHLMRR08}]
\label{thelemma}
Let $R \in {\cal R}$ such that $\mbox{Max}(R)\not=\emptyset$. Then for
all $X \in {\cal R}$ such that $X \cap R \neq \emptyset$ and $ |R|
\leq |X| \leq |\mbox{Max}(R)|$, $X$ overlaps $R$ or $\mbox{Max}(R).$
\end{lemma}

The trick we propose below for computing the overlap order of each
overlap class is also based on lemma \ref{thelemma}.

Let us assume first that we already computed all $\mbox{Max}(R).$ For
each column $c \in {\cal C}$ we compute the list $SL(c)$ of all sets
$R \in {\cal R}$ to which $c$ belongs. This list is sorted in
increasing order of the sizes of the sets respecting $\mbox{LR}$, thus
in decreasing order in $\mbox{LR}.$ Computing and sorting all lists
for all $c\in {\cal C}$ can be done in $O(|{\cal R}|)$ time using a
stable bucket sort.

Dahlhaus's overlap class identification is built on those lists. For
all $c \in {\cal C}$, let $R$ be a set containing $c$ such that
$\mbox{Max}(R)\not=\emptyset$. We define a new interval on $SL(c)$
beginning in $R$, continuing from $R$ in the order of $SL(c)$ and
finishing by the greatest row in $SL(c)$ such that $|Y|\leq
|\mbox{Max}(R)|$. Notice that this greatest row $Y$ is not necessarily equal to
$\mbox{Max}(R).$ If it is the case, the interval is said of type $M$
(for Max included), of type $E$ (for External) otherwise. Given an
interval $I$, $\mbox{First}(I)$ is the first row of the interval, thus
the row which generates the interval.

We "bucket" sort the intervals in a table $TI[1..m]$ of $m$ entries
the following way. For an interval $I=[R_{i_1} \dots R_{i_k}]$, $I$ is
added to all $TI[i_j]$, $1\geq j \geq k.$ 

\begin{figure}[htb]
\vspace{-0.3cm}
  \centering
\includegraphics[width=11cm]{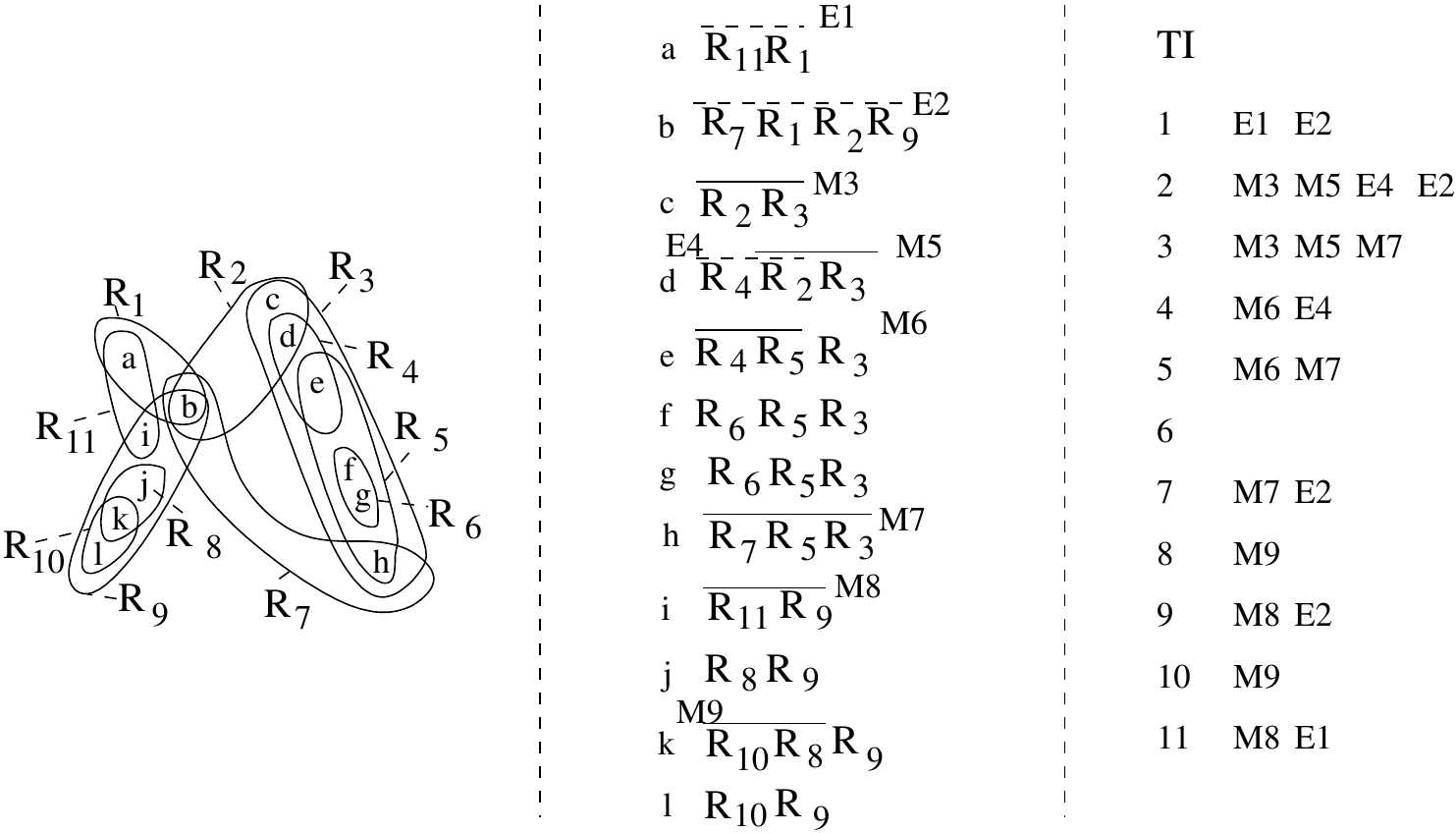}
\caption{Example: a family ${\cal R}$, its corresponding sets $SL$ and the associated $TI$ table. Intervals of type $M$ are denoted by a plain line, while intervals of type $E$ are denoted by a dash one.}
 \label{inter1}
\vspace{-0.3cm}
\end{figure}

An example of a family and the intervals associated to is shown in figure
\ref{inter1}.

To compute overlap classes, we mark them one after the other, keeping
the numbering of the overlap class each row belongs to in a table
$NC[1..m]$ all initialized to $0$.\\

\noindent
{\bf Algorithm 1: computing all overlap classes}
\begin{enumerate}
\item  Initialize the counter $nc = 1$ to count the overlap class we are tagging;
\item  Choose an arbitrary $l, 1 \leq l \leq m$ such that there exist at least on interval in $TI[l]$;
\item  For all interval(s) $I = [R_{i_1} \dots R_{i_k}],$ in  $TI[l]$.
\begin{enumerate}
\item remove all occurrences of $I$ out of $TI$;
\item mark each row in $I$ to belong to overlap class $nc$, thus $NC[i_j]=nc$, $1\leq j \leq k$;
\item recurse this algorithm from step 3 on all $i_j$, $1\leq j \leq k$, such that $TI[i_j]$ is not empty;
\item end the recursive procedure;
\end{enumerate}
\item Increment $nc$ and apply step 2 while $TI[l]$ is not empty.
\end{enumerate}

Rows that are not marked during this algorithm are themselves an overlap
class of a single element that it is not necessary to consider further
for testing C1P. We focus below on overlap classes that contain at
least 2 rows.

 By lemma \ref{thelemma}, all rows in a given interval belong to the
 same overlap class. We prove now that {\bf Algorithm 1} computes all
 overlap classes.

First, assume that 2 rows $R_i$ and $R_j$ are such that
$NC[R_i]=NC[R_j].$ Then the two rows have been marked during a
recursive call of Step 3 that recurse on each interval containing a
row. Thus the whole process computes the closure of belonging to a same
interval, which guaranties that the two rows are linked by a chain of
overlap(s).

Secondly, assume that two rows $R_1$ and $R_2$ overlap. Let us
consider wlog that $R_2 <_{LR} R_1.$ Then $\mbox{Max}(R_1)$ exists
and as $R_1$ and $R_2$ intersect on at least one column $c$, $R_2$ is
in an interval beginning in $R_1$ on $SL(c)$. We thus proved that:

\begin{proposition}[\cite{Dahlhaus00}]
Algorithm 1 computes all overlap classes of ${\cal R}.$ 
\end{proposition}

\noindent
{\em Worst case complexity of Algorithm 1}.  Algorithm 1 can be
implemented to run in $O(|{\cal R}|)$, provided that for a given row
$R$ computing $\mbox{Max}(R)$ is $O(1)$ time (see Appendix
\ref{appendix1} for details on this computation).\\

Up to now we dispose of a general scheme for computing all overlap
classes of ${\cal R}$ that is directly adapted from
\cite{Dahlhaus00,CharbitHLMRR08}. We now modify this approach to
consider the two types $M$ and $E$ of intervals successively for each row,
beginning with intervals of type $M$ and then intervals of type $E$.\\

\noindent
{\bf Algorithm 2: the computation of all overlap classes revisited}
\begin{enumerate}
\item  Initialize the counter $nc = 1$ to count the overlap class we are tagging;
\item  Choose an arbitrary $l, 1 \leq l \leq m$ such that there exist at least on interval in $TI[l]$ of type $M$;
\item  For all interval(s) $I = [R_{i_1} \dots R_{i_k}]$ of type $M$ in  $TI[l]$,
\begin{enumerate}
\item remove all occurrences of $I$ out of $TI$;
\item mark each row in $I$ to belong to overlap class $nc$, thus $NC[i_j]=nc$, $1\leq j \leq k$;
\item recurse this algorithm from step 3 on all $i_j$, $1\leq j \leq k$, such that $TI[i_j]$ is not empty;
\end{enumerate}
\item  For all interval(s) $J = [R_{i_1} \dots R_{i_k}]$ of type $E$ in  $TI[l]$,
\begin{enumerate}
\item remove all occurrences of $J$ out of $TI$;
\item mark each row in $J$ to belong to overlap class $nc$, thus $NC[i_j]=nc$, $1\leq j \leq k$;
\item recurse this algorithm from step 3 on all $i_j$, $1\leq j \leq k$, such that $TI[i_j]$ is not empty;
\item end the recursive procedure;
\end{enumerate}
\item Increment $nc$ and apply step 2 while $TI[l]$ is not empty.
\end{enumerate}

Algorithm 2 is still valid since (a) it is a simple modification of
Algorithm 1 only considering two types of intervals and (2) in each
overlap class there exist at least one interval of type $M$ to begin with at step 2.

\section{Swap Overlap Order}
\label{swap}

A swap overlap order is an order $R_{i_1} \dots R_{i_k}$ on the rows
of an overlap class such that, for all $ 2 \leq l \leq k$, at least
one of the two following cases is true:
\begin{itemize} 
\item $R_{i_l}$ overlaps one $R_{i_g}, 1 \leq g < l$,
\item $l < k$ and $R_{i_{l+1}}$ overlaps $ R_{i_g}, 1 \leq g < l$, and $R_{i_l}$ overlaps $R_{i_{l+1}}.$
\end{itemize}

\noindent
We now modify Algorithm 2 to output for each overlap class a
swap overlap order.\\

\noindent
{\bf Algorithm 3: outputing a swap overlap order for all overlap classes}
\begin{enumerate}
\item  Initialize the counter $nc = 1$ to count the overlap class we are tagging; Initialize $O_{nc}$ to the empty word $\epsilon$,
\item  Choose an arbitrary $l, 1 \leq l \leq m$ such that there exist at least on interval in $TI[l]$ of type $M$;
\item  For all interval(s) $I = [R_{i_1} \dots R_{i_k}]$ of type $M$ in  $TI[l]$,
\begin{enumerate}
\item remove all occurrences of $I$ out of $TI$;
\item concatenate to $O_{nc}$ successively the rows  $R_{i_1}$,$R_{i_k}$, $R_{i_2}$, .. ,$R_{i_{k-1}}$ 
in this order, adding a row only if $NC[i_j] =0$. After adding a row, change $NC[i_j]$ to $no$.
\item recurse this algorithm from step 3 on all $i_j$, $1\leq j \leq k$, such that $TI[i_j]$ is not empty;
\end{enumerate}
\item  For all interval(s) $J = [R_{i_1} \dots R_{i_k}]$ of type $E$ in  $TI[l]$,
\begin{enumerate}
\item remove all occurrences of $J$ out of $TI$;
\item recurse step 3 on $TI[i_1]$;
\item concatenate to $O_{nc}$ successively the rows  $R_{i_2}$, $R_{i_3}$, .. ,$R_{i_{k}}$ 
in this order, adding a row only if $NC[i_j] =0$. After adding a row, change $NC[i_j]$ to $no$.
\item recurse this algorithm from step 3 on all $i_j$, $1 < j \leq k$, such that $TI[i_j]$ is not empty;
\item end the recursive procedure;
\end{enumerate}
\item Increment $nc$ and apply step 2 while $TI[l]$ is not empty.
\end{enumerate}

The main difference with Algorithm 2 in terms of recursive call is
step 4.(b), where we first recurse on $\mbox{First}(J)$ when
considering an interval of type $E$ before processing the interval
itself.

A trace of the execution of Algorithm 3 is given in Appendix
\ref{Example1}. For the largest overlap class of our current example,
it returns the swap overlap order $O_{1}=R_2R_3R_4R_5R_7R_1R_9R_{11}.$

What is the idea behind algorithm 3 ? We begin an order by considering
and interval of type $M$, say $I = [R_{i_1} \dots R_{i_k}].$ By
placing $R_{i_1}$ and then $R_{i_k} = \mbox{Max}(R_{i_1})$ before all
other rows in $I$, Lemma \ref{thelemma} guaranties that the following
rows in $I$ overlap either $R_{i_1}$ or $R_{i_k}$.

Then, assume that there exits a row $X$ between $R_{i_1}$ and
$R_{i_k}$ in $I$. We recurse on $X$. If the line corresponding to $X$
in $TI$ contains and interval, say $I' = [R'_{i_1} \dots R'_{i_{k'}}] $,
it be of two types, $M$ or $E$.\\

\noindent
{\em Case 1}. If $I'$ is of type $M$ then it will be process first
before all type $E$ intervals corresponding to $X$. Then, either $X$
is the fist row of the interval, either not. Whatever, as $X$ already
appears in $O_{nc}$ by interval $I$, then by concatenating the rows in
the order $R'_{i_1} R'_{i_{k'}} ...$ if not already in $O_{nc}$, we
guaranty that:
\begin{itemize}
\item one of $R'_{i_{k'}} = \mbox{Max}(R'_{i_1})$ or $R'_{i_1}$
  overlaps $X$ that is already placed in $O_{nc}$ by Lemma
  \ref{thelemma}.
\item each following row in $I'$, if any, either overlaps
  $R'_{i_{k'}}$ or $R'_{i_1}$, or already appears in $O_{nc}.$\\
\end{itemize} 

\noindent
{\em Case 2}. If $I'$ is of type $E$, then $R'_{i_k'}$ is not
$\mbox{Max}(R'_{i_1}).$ Thus there is not guaranty that
$\mbox{Max}(R'_{i_1})$ (that has to exist since $I'$ is an interval
beginning in $R'_{i_1}$) has already been placed in $O_{nc}$. Thus we
first recurse on $R'_{i_1}$ (step 4-(a)) to guaranty that after some
recursion the rows $R'_{i_1}$ and $\mbox{Max}(R'_{i_1})$ appear
somewhere in $O_{nc}$ before processing $I$. Then, by lemma
\ref{thelemma}, each row following $R'_{i_1}$ in $I'$ overlaps either
$\mbox{Max}(R'_{i_1})$ or $R'_{i_1}$. As both are already in $O_{nc}$,
we simply concatenate them to $O_{nc}$ in step 4-(c).\\

Thus, summarizing the 2 cases, when concatenating new rows to $O_{nc}$,
we can insure that either (a) we add a couple $(X,\mbox{Max}(X))$,
provided that at least one of those rows overlaps a row $Y$ already placed in
$O_{nc}$ (note that if one of those rows is already in $O_{nc}$, then
the result also holds), or (b) a row $X$ that surely overlaps a row
already in $O_{nc}.$ Using this approach we identify each overlap
class and in the same time we build a swap overlap order for each
overlap class.\\

\noindent
{\em Complexity.} It is obvious that the time complexity is the same that
Algorithm 1 or Algorithm 2, that is, $O(|{\cal R|})$.

\section{Partitioning Each Overlap Class}
\label{partition-refine}

At this point, we built a swap overlap order for each non trivial
overlap class. It remains to explain how to test C1P on each such class
using this order.

We use a partitioning that is relatively similar to that of
\cite{McConnell04}, except that instead of being driven by a spanning
tree it uses a swap overlap order that is easier to build since it is
in the direct continuation of Dahlhaus's approach for computing overlap
classes. However, the important difference is that using a swap
overlap order we can not certify that we {\em cut} each time the
current partition when refined by a new row. Instead, we can certify
that if the new row $R_1$ does not cut, the following row $R_2$ will,
and $R_1$ will then cut $R_2.$ We thus {\em swap} the two rows in the
partitioning.

Let us enter details. We maintain an ordered set of sets, called {\em
  parts}, of columns of ${\cal C}.$ When adding a row, a part $C$ can
only be cut in two parts $C'C"$ such that $C'\cup C" = C$ and $C'
\cap C" = \emptyset.$ In the partitioning, $C$ is replaced by $C'C''$
or $C"C'$ depending the case, but the general order of the initial
partition is maintained.

To begin the partitioning phase, we consider the first row $R_{i_1}$
of the overlap order $O_{nc}= R_{i_1} R_{i_2} \ldots R_{i_k}$ of
overlap class $nc$. We create a first part in our partition $P_1$ that
is composed of the columns of $R_{i_1}$. We then refine this partition with
$R_2$ by first marking all elements of $R_2$. Suppose first that $R_2$
overlaps (or {\em cuts}) $R_1$ and let $X = R_1 \cap R_2$. We
partition $P$ by $R_2$ in $P_2 = (R_1\setminus X)(X)(R_2\setminus X)$,
thus we simply placed all common elements of $R_1$ and $R_2$ on a line
in such a way that both $R_1$ and $R_2$ are intervals of $P$, which is
the core of the C1P.

Let us now consider a new row $R_{i_j}$. We mark elements of
$R_{i_j}$ in $P_{j-1}$. Suppose again that $R_{i_j}$ cuts a row
already integrated to $P_{j-1}$. Let $Y$ be the set of elements of
$R_{i_j}$ that already appear in $P_j.$ Two cases may occur: 
\begin{itemize}
\item[(a)] if $Y = R_{i_j}$, we only try to group together the elements of
  $R_3$ in $P_2$. If we can,  we only cut the
  parts accordingly to build  $P_{j}$
\item[(b)] if $Y \neq R_{i_j}$, then we try to cluster the elements of $Y$
  on a border (left or right) of $P_{j-1}$. If we can, we cut the
  parts accordingly and add a new part $(R_{i_j}\setminus Y)$ before
  (resp. after) all parts of $P_{j-1}$ if the border was the left
  (resp. right) one to eventually build $P_{j}.$
\end{itemize}

\noindent
Example of partitioning on the first overlap class of our current data
set with the order $R_2R_3R_4R_5R_7R_3R_1R_9R_{11}$.
\begin{center}
\begin{tabular}{l|l|c}
Row   & Columns & Partition\\
\hline
$R_2$  & $\{b,c,d\}$ & (bcd)\\
$R_3$   & $\{c,d,e,f,g,h\}$ & (b)(cd)(efgh)\\
$R_4$   & $\{d,e\}$ & (b)(c)(d)(e)(fgh)\\
$R_5$   & $\{e,f,g,h\}$ &  (b)(c)(d)(e)(fgh)\\
$R_7$   & $\{b,h\}$&  fail \\
\end{tabular}
\end{center}

The main point of this approach is that if this process fails for a given
row, the overlap call does not verify C1P. 
 
\begin{proposition}[\cite{McConnell04}]
\label{partition2}
Let $R_{i_1} R_{i_2} \ldots R_{i_k}$ be a total order of the rows of a
given overlap class $nc$ such that each row $R_{i_j}, j> 2$, overlaps
a previous row $R_{i_l}, 1 \leq l < j$. Then the above partitioning
fails if and only if the overlap class $nc$ does not verify C1P.
\end{proposition}
\begin{preuve} 
The intuition behind this theorem is that if two rows $R_a$ and $R_b$
overlap, the intersection $X = R_a \cap R_b$ must rely in between and
the only two possible column orders respecting C1P are $(R_a\setminus
X)(X)(R_b\setminus X)$ or $(R_b\setminus X)(X)(R_a\setminus X).$

Each part of the partition derives from the intersection of two rows
or the difference of a row and its intersection with the other
rows. Thus the order of the elements inside a part is not relevant and
can be changed, but the global order of all parts is fixed and can not
be changed (not considering a global reversal) without breaking the
C1P of the previous rows. This has for consequence that when adding a
new row that overlaps (at least) one row that is already embedded in
the current permutation, C1P will be maintained only if the elements
of the new rows can be embedded in $P$ respecting the order of its
parts. The fact that the order of the elements inside each parts is
not relevant allows us to split some parts (placed in the extremities
of the touched zone) in two subparts, those touched by the new row on
a side, the rest on the other side. This is the only operation
authorized when adding a row to test if we can maintain C1P adding the
new row.

A new row can be embedded in $P$ under those conditions only in the
two cases (a) and (b) equivocated above. Therefore, if the partitioning is
feasible, then the new partition ``encodes'' all possible column order for
the set of rows considered up to this point to verify C1P. If not, this insures that
no column order could be valid for the set of rows to verify C1P.
\end{preuve}

In our approach, as we manipulate swap overlap orders, the
partitioning phase must be slightly modified in the following
way. Suppose that we want to refine the partition $P_{j-1}$ with
$R_{i_j}$. If $R_{i_j}$ does not overlap any previous row use in the
partitioning, that is if all columns of $R_{i_j}$ either belong to
the same part of $P_{j-1}$ of to none, we swap $R_{i_j}$ and
$R_{i_{j+1}}$, refine the partition with $R_{i_{j+1}}$ and only then
with $R_{i_j}.$ The swap overlap order guaranties that $R_{i_{j+1}}$
will cut a previous row, and that $R_{i_j}$ overlaps $R_{i_{j+1}}.$ We
call this partitioning a {\em swap partitioning.}

\begin{theorem}
\label{partition}
Let $R_{i_1} R_{i_2} \ldots R_{i_k}$ be a swap overlap order of the
rows of a given overlap class $nc.$ Then the above swap partitioning
fails if and only if the overlap class $nc$ does not verify C1P.
\end{theorem}
\begin{preuve}
By swapping the rows when necessary, we insure that the order of the
$R_{i_1} R_{i_2} \ldots R_{i_k}$ rows in which we refine the partition
verifies that each row $R_{i_j}, j> 2$, overlaps a previous row
$R_{i_l}, 1 \leq l < j$, thus satisfying the conditions of proposition
\ref{partition2}.
\end{preuve}

\noindent
{\bf Implementation issues.} Let us now consider the time complexity
of our partitioning. We show below how it might be implemented in time
$O(|O_{nc}|)$ where $|O_{nc}|$ is the sum of the size of all rows
belonging to the overlap class.

\noindent
The data structure we need must allow us to
\begin{enumerate}
\item split a part $C$ in $C'C''$ in the number of the elements of $C$ touched;
\item add a new part to the left of to the right of the current
partition in the number of the elements added;
\item test if the elements touched can be made consecutive;
\item test if a new row cut another one already embedded in the partition;
\end{enumerate}

There might be many data structures implementation having these
properties. We propose below a simple one. This structure can also
replace that used in \cite{CharbitHLMRR08} for identifying all
$\mbox{Max}(X)$ used by Dahlhaus's algorithm (see Appendix
\ref{appendix1}), and thus our whole algorithm only uses a single data
structure.

\begin{figure}[htb]
\centering \includegraphics[width=7cm]{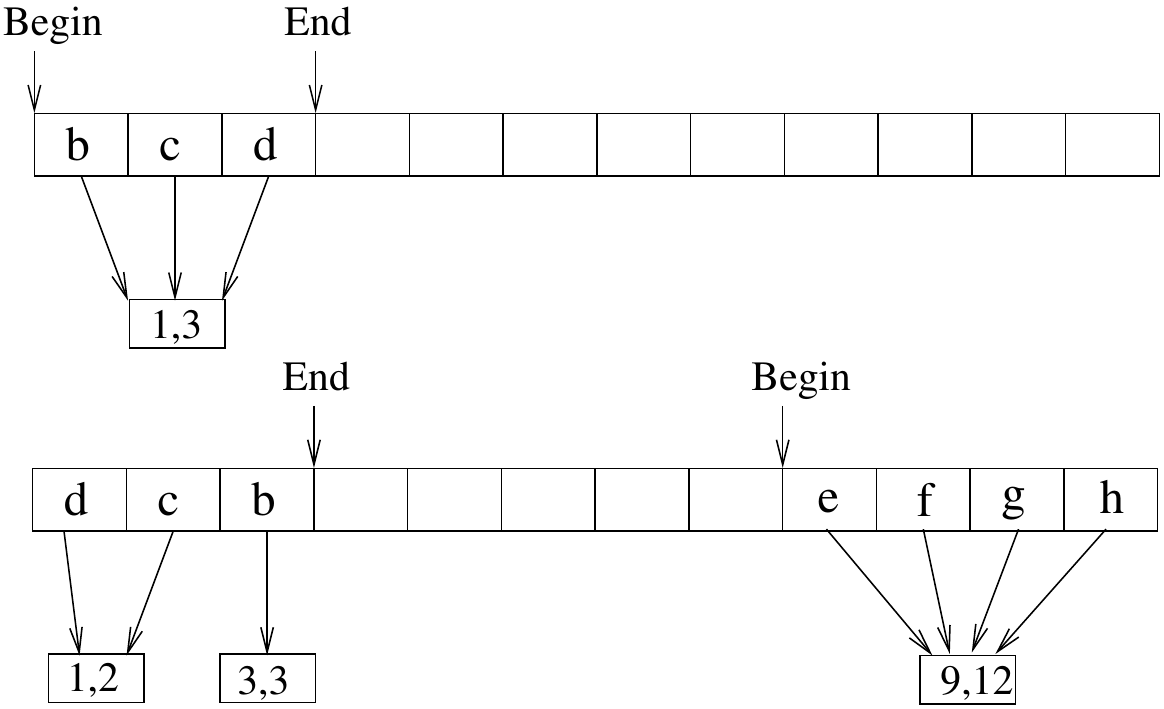}
\caption{Example continued: implementation of
$(bcd)$ and then $(efgh)(dc)(b)$ when refining $R_2 = \{b,c,d\}$ by $R_3 = \{c,d,e,f,g,h\}.$}
 \label{lcapic3}
\end{figure}

We basically use an array of size $|{\cal C}|$ to store a stack which
encodes a permutation of elements of $C$. Each cell of this array
contains a column and a link to the part it belongs to. A part is
coded as a pair of its beginning and ending positions in the array,
relatively to the beginning of the array. A schematic representation
of this data structure is given in Figure \ref{lcapic3}.

\noindent
Using this data structure, refining a part $C$ by one of its subset
$C''$ can be easily done in $O(|C''|).$ Indeed, let $[i,j]$ be the
bounds of $C$. We swap elements in the subtable $[i,j]$ to place all
$s= |C''|$ elements of $C''$ at the end or at the beginning of this
subtable as necessary. We then adjust the bounds of $C$ to $[i,j-s]$
or $[i+s,j]$ depending of the case and create a new set $[j-s+1,j]$
or $[i,i+s-1]$ on which the $s$ elements of $C''$ now point.

Adding a new part to the left of to the right of the current partition
in the number of the elements added is easy since it suffices to
create a new part and move the pointers of the beginning or ending
modulo $|{\cal C}|.$ An example of such operation is shown in Figure
\ref{lcapic3}.

Assume that a new row $R$ used for refining cut a class in the
partition $P$, and let $Y \subset R$ be the elements of $R$ that are
already in the partition.

If $Y \neq R$, then, to verify C1P, all classes touched by $Y$ must be
placed at an extremity of $P$, all parts from this extremity must be
fully touched except the last one of which all elements touched has to
be placed on the side of the extremity we considered. All these
requirements can easily be checked in the number of elements of $R$,
and if they are verified, a new part containing $R \setminus Y$ is
added to the extremity.

If $Y = R$, then to verify C1P there should be a left part that might
not be fully touched followed by a series (that can be empty) of plenty
touched parts and eventually a last part also not necessary fully
touched. This is also not difficult to check in $O(|R|).$

The novelty in our approach is that a new row $R$ might not cut the
current partition, which has to be tested efficiently. This can also
easily be checked in $O(|R|)$ on our structure. Indeed, it suffices to
test if $R$ is included in a single part, in none, or contains all
parts. We thus have:

\begin{theorem}
\label{C1Poverlapclass}
Testing the C1P of the rows belonging to a same overlap class can be
done in $O(|O_{nc}|)$ time provided a swap overlap order $O_{nc}$ of it.
\end{theorem}

\noindent
And eventually:

\begin{corollary}
Testing the C1P of a family ${\cal R}$ can be done in $O(|{\cal R}|)$
using a swap overlap order of each overlap class.
\end{corollary}
\begin{preuve}
It suffices to compute all overlap classes of ${\cal R}$ using
Algorithm 3 that provides for each overlap class a swap overlap
order. Then Theorem \ref{C1Poverlapclass} insures that C1P can be
tested on each overlap class in the number of rows belonging to this
class. As overlap classes partition ${\cal R}$ and that ${\cal R}$
verifies C1P if an only if each overlap class verifies C1P, the whole
test can be done in $O(|{\cal R}|)$ time.
\end{preuve}

\appendix
\newpage
\section{Trace of Algorithm 3 on Our Example}
\label{Example1}

We trace below the recursive steps of Algorithm 3 on our example for
the identification of the first overlap class while outputting the
order $O_{1}=R_2R_3R_4R_5R_7$ $R_3R_1R_9R_{11}.$

{\small 
\begin{multicols}{2}
\begin{enumerate}
\item Step 1. $O_{nc}=\varepsilon$, $nc= 1$
\item $TI[2]$ is choosen in step $2$ since it contains an interval of type $M.$
\item Step $3_1$. We consider $I_1= M3 = [R_2R_3]$    
\item $3_1$-(a) all occurrences of $M3$ are removed out of $TI$
\item $3_1$-(b) $O_{1}=R_2R_3$, $NC[2] = 1$, $NC[3] = 1$
\item  $3_1$-(c) Recursive call to Step $3_2$ on $R_2$ from $M3$. We consider $I_2= M5 = [R_2R_3]$ 
\item $3_2$-(a) all occurrences of $M5$ are removed out of $TI$
\item $3_2$-(b) as $NC[2] = NC[3] = 1$ no row si added to $O_{1}$
 \item $3_2$-(c) Recursive call to Step $3_3$ on $R_2$ from $M5$. 
\item Entering Step $4_3$ since there is no more interval of type $M$ in $TI[2]$. We consider $I_3= E4 = [R_4R_2]$ 
\item $4_3$-(a) all occurrences of $E4$ are removed out of $TI$  
\item $4_3$-(b) Recursive call to Step $3_4$ on $TI[4]$. We consider $I_4= M6 = [R_4R_5]$  
\item $3_4$-(a) all occurrences of $M6$ are removed out of $TI$  
\item $3_4$-(b) $O_{1}=R_2R_3R_4R_5$, $NC[4] = 1$, $NC[5] = 1$
\item $3_4$-(c) Recursive call to Step $3_5$ on $TI[4]$. As $TI[4]$ is now empty, we return to step $3_4$
\item  $3_4$-(c) Recursive call to Step $3_6$ on $TI[5]$. We consider $I_6= M7 = [R_7R_5R_3]$ 
\item  $3_6$-(a) all occurrences of $M7$ are removed out of $TI$
\item  $3_6$-(b) $O_{1}=R_2R_3R_4R_5R_7$, $NC[7] = 1$
\item  $3_6$-(c) Recursive call to Step $3_7$ on $R_7$ from $M7$. We consider $I_7= E2 = [R_7R_1R_2R_9]$ 
\item  $4_7$-(a) all occurrences of $E2$ are removed out of $TI$
\item $4_7$-(b) Recursive call to Step $3_8$ on $TI[7]$. As $TI[7]$ is now empty, we return to step $4_7$
\item $4_7$-(c) $O_{1}=R_2R_3R_4R_5R_7R_1R_9$, $NC[1] = 1$, $NC[9] = 1$
\item $4_7$-(d) Recursive call to Step $3_9$ on $R_1$ We consider $I_9= E1 = [R_{11}R_1]$ 
\item  $4_9$-(a) all occurrences of $E1$ are removed out of $TI$
\item  $4_9$-(b) Recursive call to Step $3_{10}$ on $R_{11}$ We consider $I_{10}= M8 = [R_{11}R_9]$
\item  $3_{10}$-(a) all occurrences of $M8$ are removed out of $TI$
\item  $3_{10}$-(b) $O_{1}=R_2R_3R_4R_5R_7R_1R_9R_{11}$, $NC[11] = 1$
\item  $3_{10}$-(c) Recursive call to Step $3_{11}$ on $TI[11]$. As $TI[11]$ is now empty, we return to step $3_{10}$
\item  $3_{10}$-(c) Recursive call to Step $3_{12}$ on $TI[9]$. As $TI[9]$ is now empty, we return to step $3_{10}$ than also ends, returning to Step $4_9$
\item  $4_9$-(c) Nothing to concatenate from $E1 = [R_{11}R_1]$ since the two rows are already in  $O_{1}$.
\item  $4_9$-(d)  Recursive call to Step $3_{13}$ on $TI[1]$. As $TI[1]$ is now empty, we return to step $4_{9}$ which also ends, thus returning to Step $4_7$-(d)
\item $4_7$-(d) Recursive call to Step $3_{14}$ on $TI[2]$.  As $TI[2]$ is now empty, we return to step $4_{7}$-(d)
\item $4_7$-(d) Recursive call to Step $3_{15}$ on $TI[9]$.  As $TI[9]$ is now empty, we return to step $4_{7}$ which also ends, thus returning to Step  $3_6$-(c)
\item $3_6$-(c) Recursive call to Step $3_{16}$ on $R_1$ from $M7$.  As $TI[1]$ is now empty, we return to step $3_6$-(c)
\item  $3_6$-(c) Recursive call to Step $3_{17}$ on $R_2$ from $M7$.  As $TI[2]$ is now empty, we return to step $3_6$-(c)
\item $3_6$-(c) Recursive call to Step $3_{18}$ on $R_9$ from $M7$. As $TI[9]$ is now empty, we return to step $3_6$-(c) which also ends, returning to Step $3_4$-(c)
\item $3_4$-(c) Recursive call to Step $3_{19}$ on $R_3$ from $M7$. As $TI[3]$ is now empty, we return to step $3_4$-(c) which also ends, returning to Step  $4_3$-(b)
\item $4_3$-(c) Recursive call to Step $3_{20}$ on $R_2$ from $E4$.  As $TI[2]$ is now empty, we return to step $3_4$-(c) which also ends, returning to Step $3_2$-(c).
\item $3_2$-(c) Recursive call to Step $3_{21}$ on $R_3$ from $M5$.  As $TI[3]$ is now empty, we return to step $3_2$-(c) which also ends, returning to Step $3_1$-(c).
\item $3_1$-(c) Recursive call to Step $3_{22}$ on $R_3$ from $M3$.   As $TI[3]$ is now empty, we return to step $3_1$-(c) which also ends, ending the identification fo the first overlap class. The returning order for $nc=1$ is thus $O_{1}=R_2R_3R_4R_5R_7R_1R_9R_{11}.$
\end{enumerate}
\end{multicols}}

\section{Computing all $\mbox{Max}(X)$}
\label{appendix1}

In this appendix we recall the computation of $\mbox{Max}(R)$ only
slightly modified compared to the that published in
\cite{CharbitHLMRR08}. The very small modifications is that we impose
$\mbox{Max}(R)$ to be greater or equal to $R$ in the $LR$ order, while
in \cite{CharbitHLMRR08} the constraint for $\mbox{Max}(R)$ is only to
be of size greater or equal to that of $R$. This implies that in
\cite{CharbitHLMRR08} and also in the original paper of Dahlhaus
\cite{Dahlhaus00} $\mbox{Max}(R)$ can be after $R$ in the $LR$ order
if $|\mbox{Max}(R)|=|R|,$ which in fact complexifies the understanding
of the algorithm.

 We consider a boolean matrix $\mbox{BM}$ of size $|{\cal F}|\times
 |{\cal C}|$ such that each row represents a set $R\in {\cal F}$ in
 the order of $\mbox{LR}$, and each column an element $c \in {\cal
   C}.$ The value $\mbox{BM}[i,j]$ is $1$ if and only if $c_j \in
 R_i.$

\begin{figure}[htb]
\vspace{-0.3cm}
  \centering
\includegraphics[width=7cm]{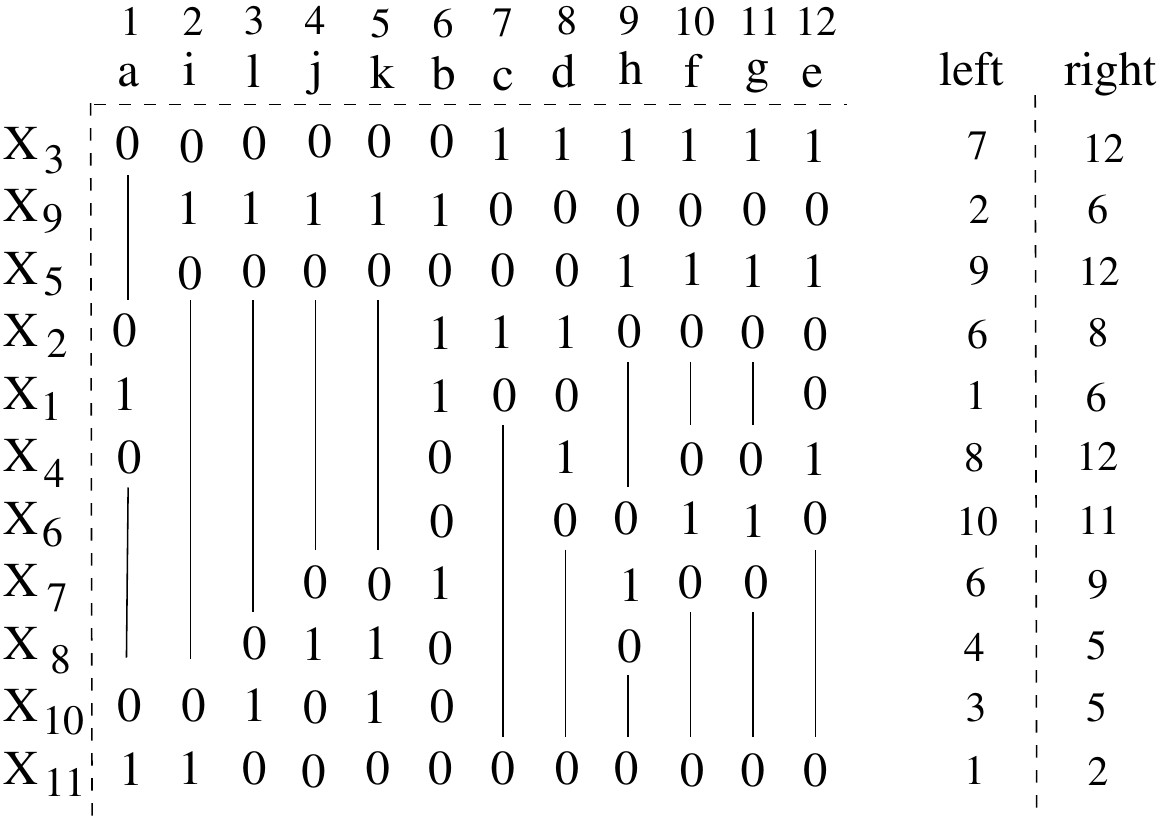}
\caption{Example continued: $BM$ matrix which lines are sorted in $LR$ order and which columns are sorted in lexicographic order.}
 \label{matrix}
\vspace{-0.3cm}
\end{figure}

Let us consider first below that all columns of $BM$ are
lexicographically sorted. Figure \ref{matrix} shows the $BM$ matrix
for the set family of Figure \ref{inter1}.

For each $R \in {\cal F}$ we denote $\mbox{left}(R)$
(resp. $\mbox{right}(R)$) the number of the column of $BM$ containing the
leftmost (resp. rightmost) $1$ in the row of $R$.

\begin{lemma}[\cite{CharbitHLMRR08}]
Let $R_1,R_2\in {\cal F}$ such that $R_2$ overlaps $R_1$ in $BM$. Then there exists a row $R \leq_{LR} R_2$ such that $\mbox{BM}[R,\mbox{left}(R_1)]=0$ and
 $\mbox{BM}[R,\mbox{right}(R_1)]=1.$
\label{goodlemma}
\end{lemma}

\begin{lemma}[\cite{CharbitHLMRR08}]
 Let $R_1\in {\cal F}.$ Then $\mbox{Max}(R_1)\not=\emptyset$ if and only if
 there exists a row $R$ in $BM$ such that
 $\mbox{BM}[R,\mbox{left}(R_1)]=0$ and $\mbox{BM}[R,\mbox{right}(R_1)]=1$
 and verifying $|R| \geq_{LR} |R_1|$. 
\label{uplemma}
\end{lemma}

\begin{lemma}[\cite{CharbitHLMRR08,Dahlhaus00}]
Let $R_1\in {\cal F}$ such that $\mbox{Max}(R_1)\not=\emptyset.$ Then
$\mbox{Max}(R_1)$ corresponds to the highest row $R$ in $BM$ such that
$\mbox{BM}[R,\mbox{left}(R_1)]=0$ and $\mbox{BM}[R,$ $\mbox{right}(R_1)]=1.$
\label{lemup}
\end{lemma}

Dahlhaus's approach for computing all $\mbox{Max}(R_1)$ the smallest
$R$ in $LR$ order such that $\mbox{BM}[R,\mbox{left}(_1)]=0$ and
$\mbox{BM}[R,\mbox{right}(R_1)]$ $=1.$ Dahlhaus reduces the problem to
LCA computations, which has been simplified in \cite{CharbitHLMRR08}
using partitions.\\

\noindent
{\bf Computing all $\mbox{Max}(R)$ using set partitioning.} We
manipulate sorted partitions of $V$ that we refine by each $R \in
{\cal R}$ taken in $\mbox{LR}$ order, that is, in decreasing order of
their sizes. The initial partition is the whole set ${\cal C}$ and
denoted $P_{\cal C}$. The refinement is slightly restricted compared
to that of Section \ref{partition-refine} in the sense that $C$ is
always split in $C'C''$ (and never $C''C'$) if $C''$ represents the
set of elements in $R.$ Refining a partition $P$ by a set $R \in {\cal
R}$ consists in refining successively all parts in $P$. We note this
refinement $P|_R.$

For example (continued), if
$P=\{a\}\{i,j,k,l\}\{b\}\{c,d\}\{e,f,g,h\}$ and $R = R_4 =
\{d,e\}$, $P|_R = \{a\}\{i,j,k,l\}\{b\}\{c\}\{d\}\{f,g,h\}\{e\}.$\\

\noindent
The approach requires 3 steps:
\begin{enumerate}
\item refine $P_V$ by all $R \in {\cal R}$ taken in
  $\mbox{LR}$ order;
\item then compute for each $R \in {\cal R}$ the values of
  $\mbox{left}(R)$ and $\mbox{right}(R)$ and sort all $R \in {\cal
  R}$ in a special order in regard with these values;
\item eventually refine $P_V$ again by all $R \in {\cal R}$
taken in $\mbox{LR}$ order but using the informations computed in step
2 to compute all $\mbox{Max}(R).$
\end{enumerate}
\noindent
These 3 steps are detailed below. 

\paragraph{\bf \em Step 1 \-- Refining $P_V.$}

Let us consider the final partition we obtain after refining $P_V$ by
each $R \in {\cal R}$ taken in $LR$ order. We note this partition
$P_f$.

\begin{lemma}[\cite{CharbitHLMRR08}]
The elements of $P_f$ are sorted accordingly to the lexicographical order
of the columns of $BM.$
\label{lexico}
\end{lemma}

\noindent
For example (continued), on the data in Figure \ref{matrix},
  $P_f=\{a\}\{i\}\{l\}\{j\}\{k\}$$\{b\}$ $\{c\}$$\{d\}\{h\}\{f,g\}\{e\}.$
  Note that equal columns of $BM$ are in the same part of $P_f$ on
  which we fix an arbitrary order.

\paragraph{\bf \em Step 2 \-- Computing all $\mbox{left}(R)$ and $\mbox{right}(R)$ values.}

We then compute all $\mbox{left}(R)$ and $\mbox{right}(R)$ values on
$P_f.$ This can be done easily in $O(|{\cal R}|+n)$ time by scanning each
$R \in {\cal R}$ and keeping the minimum and maximum position of one
of its element in $P_f$. We also compute a data structure $AM$ that
for each position $1 \leq i \leq |V|$ of $P_f$ gives a list of all $R
\in {\cal R}$ such that $i=\mbox{right}(R)$. All those lists are sorted in
increasing order of $\mbox{left}(R).$ The structure also allows an
element $R \in {\cal R}$ to be removed from the list
$AM[\mbox{right}(R)]$ in $O(1)$ time. This can be insured for instance
using doubly linked list to implement each list, and the whole
structure can easily be built in $O(n+m)$ time using bucket sorting.

\paragraph{\bf \em Step 3 \-- Refining $P_V$ again and identifying all $\mbox{Max}(R).$}

The main idea is the following. Assume that at a step of the
refinement process in $LR$ order we refine a part $C=\{c_{i_1},\ldots,
c_{i_k}\}$ of a partition $P$ by $R_2\in {\cal R}$ and that it results two
non empty parts $C'C''.$

\begin{lemma}[\cite{CharbitHLMRR08}]
Let $R \in {\cal R}$ such that $|R| \leq |Y_2|$,  $\mbox{left}(R) \in C'$
and $\mbox{right}(R)\in C''.$ Then $R_2=\mbox{Max}(R).$
\label{assignmax}
\end{lemma}

\noindent
The last phase of the algorithm thus consists in refining $P_{\cal C}$ again by
all $R_2 \in {\cal R}$ taken in $LR$ order. We first initialize all
values $\mbox{Max}(R)$ to $\emptyset$. Each time a new split $C'C''$
appears (say between positions $l$ and $l+1$), for all $c \in C''$ all
lists $AM[c]$ are inspected the following way: let $R$ be the top of
one of those the list; while $\mbox{left}(R)\leq l$, $R$ is popped off
the list and $\mbox{Max}(R) \leftarrow R_2$. After having refined with
$R_2$, $R_2$ is removed from the $AM$ structure.

\begin{lemma}[\cite{CharbitHLMRR08}]
The above algorithm correctly computes in 3 steps all $\mbox{Max}(R)$, $R
\in {\cal R}$.
\label{correctness}
\end{lemma}

\noindent 

The partition refinement can be efficiently implemented using the data
structure presented in Section \ref{partition-refine} of that in \cite{CharbitHLMRR08} which is
a simpler version of the first one.

\begin{theorem}[\cite{CharbitHLMRR08}]
The identification of all $\mbox{Max}(R), \; R \in {\cal R},$ using
partition refinement can be done in $\Theta(|{\cal R}|)$ time. 
\end{theorem}

\end{document}